\documentclass[12pt]{cernart}              

\usepackage{graphicx,color}       
\usepackage{epsfig,rotating}      

\newcommand{\gf}{$G_\mathrm{F}$}

\newcommand{\mupl}{$\mu^{+}$}
\newcommand{\pipl}{$\pi^{+}$}
\newcommand{\taumu}{$\tau_{\mu}$}

\newcommand{\pitomu}{$\pi \rightarrow \mu$}
\newcommand{\mutoel}{$\mu \rightarrow \mathrm{e}$}
\newcommand{\pitomutoel}{$\pi \rightarrow \mu \rightarrow \mathrm{e}$}
\newcommand{\musr}{$\mu SR$}

\newcommand{\etal}{\textit{et al.}}

\newcommand{\nph}{\textit{Nuc.\,Phys.}}

\newcommand{\prl}{\textit{Phys.\,Rev.\,Lett.}}
\newcommand{\phl}{\textit{Phys.\,Lett.}}

\begin{document}

\begin{titlepage}
\docnum{CERN-PH-EP/2007-xxx}
\date{July 21, 2007}
\title{\bf \Large Measurement of the Fermi Constant by FAST}

\begin{Authlist}
\begin{center}

A.~Barczyk, J.~Kirkby, L.~Malgeri \\
{\em \small CERN, Geneva, Switzerland} \\[2mm]

J.~Berdugo, J.~Casaus,  C.~Ma\~{n}\'{a}, J.~Marin, G.~Martinez,  E.~S\'{a}nchez, C.~Willmott \\
{\em \small CIEMAT, Madrid, Spain} \\[2mm] 

C.~Casella, M.~Pohl \\
{\em \small Universit\'e de Gen\`eve, Geneva, Switzerland} \\[2mm] 

K.~Deiters, P.~Dick, C.~Petitjean  \\
{\em \small Paul Scherrer Institute, Villigen, Switzerland} \\[5mm]

{\bf FAST Collaboration}  \\[15mm] 

\it{Submitted to Physics Letters B}

\end{center}

\end{Authlist}
\end{titlepage}

\pagenumbering{roman}  
\setcounter{page}{2}  
\newpage \mbox{} \newpage  
 
\pagenumbering{arabic}  
\setcounter{page}{1}  

\begin{abstract}   
An initial measurement of the lifetime of the positive muon to a precision of 16 parts per million (ppm) has been performed with the FAST\footnote{FAST is an acronym of \textit{Fibre Active Scintillator Target.}} detector at the Paul Scherrer Institute \cite{fast_proposal}.  The result is $\tau_\mu$ = 2.197 083 (32) (15) $\mu$s, where the first error is statistical and the second is systematic. The muon lifetime determines the Fermi constant, \gf\ $=  1.166 \; 353 (9) \times 10^{-5}$ GeV$^{-2}$ (8 ppm). 
 
\end{abstract}

\section{Introduction}
\label{sec_introduction}

The Standard Model has three free parameters in the bosonic sector: the electromagnetic coupling constant, $\alpha$, the mass of the Z boson, $m_\mathrm{Z}$, and the Fermi coupling constant, \gf. The theory becomes predictive when these and several other fundamental parameters have been determined experimentally. By progressively improving the precision of these parameters, the theoretical predictions become increasingly precise  and, in turn, the experimental measurements are increasingly sensitive to new physics beyond the Standard Model. Therefore, on quite general grounds, it is important for each of the fundamental parameters of the Standard Model to be measured with the highest possible experimental precision.

The Fermi coupling constant is determined from the measurement of the positive muon lifetime, \taumu, through the relationship
\begin{equation}
\frac{1}{\tau_\mu} = \frac{G_\mathrm{F}^2 m_\mu^5}{192\pi^3} \; (1+\Delta q).
\label{eq_gf}
\end{equation}
In order to avoid uncertainties in the capture rate of negative muons on the target nuclei, the more precise value of \gf\ is derived from the positive muon lifetime. In this equation  $\Delta q$ encapsulates the higher order QED and QCD corrections calculated in the Fermi theory, in which the weak charged current is described by a contact interaction. This is discussed in detail by van Ritbergen and Stuart \cite{ritbergen00}, who have computed the second-order QED corrections and reduced the theoretical uncertainty of the radiative corrections to 0.3 ppm. In contrast to the case for $\alpha$, \gf\  is not afflicted with a hadronic uncertainty until 2-loops is reached, and consequently its influence is negligible.

Among the other parameters in Eq.\,\ref{eq_gf}, the error on the muon mass, $m_\mu$, contributes an uncertainty of 0.21 ppm to \gf. The contribution from the uncertainty in the muon neutrino mass is negligible, assuming the mass limits implied by neutrino oscillation experiments. The largest uncertainty in \gf\ comes from the error on \taumu, which is 18 ppm in the 2006 Review of Particle Properties \cite{pdg06}, and recently reduced by a new measurement from the MuLan experiment at PSI to a precision of 11 ppm \cite{mulan07}. In this paper, an initial measurement of $\tau_{\mu}$ is presented from the FAST experiment with a precision of 16 ppm. The final goal of FAST is an uncertainty  of 2 ppm in \taumu, which will determine \gf\ to 1~ppm precision.

\section{The FAST detector}

\begin{figure}[tbp]
  \begin{center}
      \makebox{\includegraphics[width=138mm]{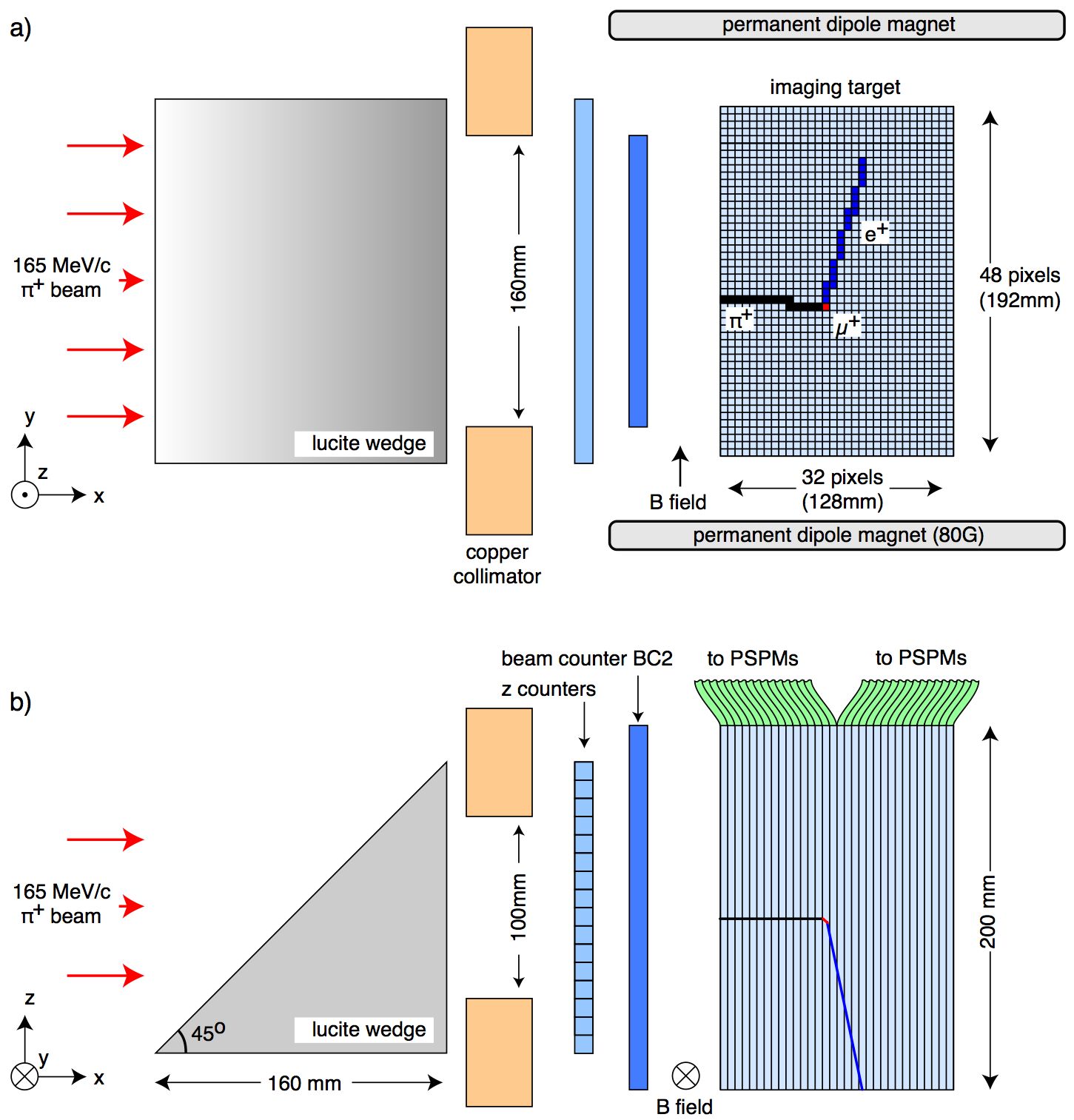}}
  \end{center}
  \caption{Schematic drawing of the FAST detector: a) top view, and b) side view. A representative event shows the \pipl\ beam particle stopping in the target followed by a \pitomu\ decay and finally a \mutoel\ decay. This sequence is imaged by the target in the $x y$ projection and the decay times recorded.}  
  \label{fig_fast_detector} 
  \end{figure}

The FAST detector (Fig.~\ref{fig_fast_detector}) operates in the $\pi$M1 secondary beamline of the 590 MeV proton cyclotron at the Paul Scherrer Institute, which generates a 2 mA DC proton beam (100\% duty cycle). 
The detector is a fast imaging target of $32 \times 48$ pixels constructed from plastic scintillator bars of dimension $4 \times 4 \times 200$~mm$^3$. The target is read out by wavelength shifter fibres attached to Hamamatsu H6568-10 position sensitive photomultipliers (PSPMs). There are 96 PSPMs in total, each one viewing $4 \times 4$ pixels. A DC $\pi^{+}$ beam of momentum 165 MeV/c is stopped in the target and the decay times of each \pitomutoel\ decay chain are registered. A wedge-shaped beam degrader ensures that the pion stopping positions are distributed uniformly through the target. A local copper beam collimator defines a beam aperture of $160 \times 100$~mm$^2$ and suppresses pion stops near the edges of the target. An array 16 finger counters (Z counters), each $10 \times 10 \times 200$~ mm$^2$, measures the vertical position of the beam particle as it enters the target.  

In order to reach the final goal of a 1 ppm measurement of \gf, FAST must handle a factor 100 higher event rate than earlier muon lifetime experiments and, at the same time, reduce the systematic errors by at least an order of magnitude. The high rate is achieved by means of a highly granular and fast detector that can measure several muon decays in parallel. The imaging target (Fig.\,\ref{fig_fast_detector}) essentially comprises a large number of replicated mini-detectors placed side-by-side, each one capable of making an independent and simultaneous measurement of a \mupl\ decay.  With regards systematic uncertainties, the detector is designed to suppress several effects that limited previous experiments. Firstly, pile-up effects are reduced by a fast, imaging detector. Secondly, possible gain shifts and additional event losses at the start of the muon decay period due to the higher counting rate are avoided by operating in a DC beam (uniform counting rate).  Thirdly, muon spin polarisation systematics are highly suppressed by a) using a \pipl\ beam (spin 0) and tagging the \pitomu\ transition, b) applying range cuts that further suppress residual beam muons, c) designing a uniform detector with large angular acceptance, and d) providing an 80~G magnetic field across the target, which causes rapid precession of the muons so that any residual spin effects can be observed and measured.

The PSPM signals are passed through pre-amplifiers into custom-designed updating discriminators with dual-threshold differential ECL outputs. The low level (LL) threshold is satisfied by minimum ionizing particles, and the LL pulses are sent to the time-to-digital converters (TDCs). The high level (HL) threshold is set to efficiently detect the large pulses from stopping pions or \pitomu\ decays (4 MeV muon kinetic energy), but to suppress minimum ionizing particles. The HL pulses are sent to the level 2 (LV2) trigger.

The time measurement is performed with 16 CAEN V767 128-channel multihit TDCs~\cite{TDC}. They provide a fast time stamp for each PSPM pulse and then load the pixel address and time into memory. Dual hit registers on each input channel of the TDCs ensure a short minimum time separation of 10 ns for two hits to be registered on the same channel, although a longer deadtime of up to several 100 ns occurs before a third hit can be registered on the same channel since the contents of the hit registers must first be written into the local event buffer. The TDCs are driven by an external 30 MHz Rb atomic clock. This defines a precise so-called \textit{coarse} TDC tick which is then divided by the delay locked loop of the TDC chip into 32 \textit{fine} TDC ticks, each of 1.041667 ns. The time window of the TDCs is set to measure decays between -8~$\mu$s and +22~$\mu$s relative to the pion stop time.

The level 1 (LV1) trigger defines an incident beam particle by a coincidence of three beam counters (BC1, BC2 and Z) with the accelerator RF signal. The leading edge of the LV1 trigger is determined by the machine RF pulse (period $\sim$19~ns). The LV1 deadtime is 50 ns, implying that no further triggers are accepted in the two subsequent  RF  buckets.

The LV2 trigger uses the HL data from the target to a) reconstruct the pion stopping pixel and b) identify the subsequent \pitomu\ decay. Having identified a prompt stopping beam pion, in-time with LV1, the LV2 trigger is satisfied by a delayed second pulse---within a 15--100 ns window---on any of the $3 \times 3$ pixels centred on the pion stop pixel. The LV2 trigger then provides a selective trigger pulse for the TDCs subtended by a $7 \times 7$ pixel array centred on the pion stop pixel. This serves to reduce the data bandwidth requirements of the data acquisition system (DAQ) since, on average, only 2.4 out of the 16 TDC modules are triggered per event. The $x$ and $y$ coordinates of the muon stop pixel are also sent to TDC data inputs for use in the subsequent event analysis. The experiment reads out only TDC data; each 128~ch TDC has 96 ch devoted to target readout and 32~ch devoted to other data, including the pion stop coordinates. The LV2 trigger operates at up to 1~MHz trigger rate and can process multiple LV1 triggers in parallel, provided they are separated by at least 50~ns.

The very high data rate of the FAST detector makes it impractical to store all events for later analysis offline. Not only would the disk space requirements be huge---about 7~TB/day---but the reading and reprocessing time for these events would far exceed the original time taken for their collection. The adopted solution is to carry out a full analysis of all data online, and to record approximately each hour a set of about 1200 separate muon lifetime histograms that contain all the information required for the lifetime measurement and for the study of systematic errors. In addition, a pre-scaled fraction of the raw events---of order 1\% depending on the rate---are recorded for monitoring purposes and to study specific aspects of the systematic errors. A web-based online monitor program accumulates control histograms to verify the proper functioning of the detector and provides a warning if any distribution falls outside its pre-defined tolerance. The monitor program includes a software emulation of the LV2 trigger (albeit with LL data) and continuously verifies the correct functioning of the LV2 hardware trigger.

The analysis procedure uses the muon coordinates given by LV2 and identifies any positron candidate emerging from the muon pixel in the -8 to 22~$\mu$s window. Positron tracks are identified using a set of 512 mask topologies in a so-called $5 \times 5$ \textit{superpixel} matrix centred on the muon pixel. The mask topologies are chosen to be highly efficient for accepting true positrons from muon decays, while suppressing  false positrons due to overlapping tracks from other beam particles or decays elsewhere in the target. The absolute positron time is taken as the average of all the positron pixels within the superpixel.

The readout of the TDC's is done via PVIC boards \cite{pvic} which provide an interface between the DAQ PCs and the 4 VME crates containing the TDC modules. The data are transferred at the full VME bandwidth limit of 20~MB/s per crate into 4 DAQ PCs. A Gigabit ethernet switch and collector PC then build the events from the DAQ PCs into time slices and pass these slices in turn to a farm of event analyser PCs. The event analyser PCs process the events and write the histograms and other information to disk.

\section{Data sample}

\begin{figure}[tbp]
  \begin{center}
      \makebox{\includegraphics[width=160mm]{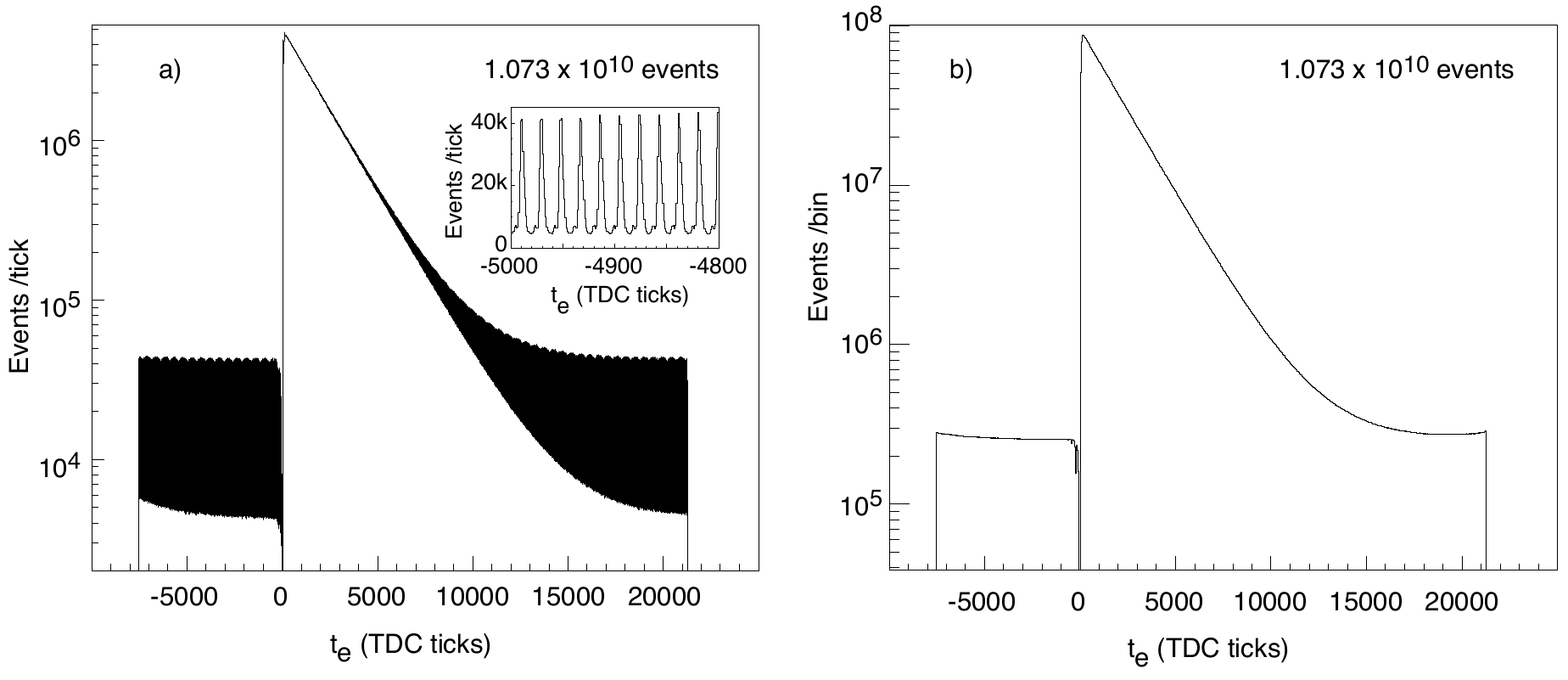}}
  \end{center}
  \caption{Muon lifetime distribution for the full sample of $1.073 \times 10^{10}$ events, with the data binned by intervals of a) a fine TDC tick (1.041667~ns), and b) an accelerator RF period ($\sim$19~ns). The electron time, t$_e$, is measured with respect to the RF bucket containing the beam pion. The insert in panel a) shows part of the negative decay region on an expanded scale, where the accelerator RF period of the background events is clearly visible. }  
  \label{fig_lifetime} 
  \end{figure}

The data were recorded in December 2006 at an average LV2 trigger rate of 30 kHz. The muon lifetime histogram for the full data sample of $1.073 \times 10^{10}$ events is shown in Fig.\,\ref{fig_lifetime}a, measured with the resolution of a fine TDC tick (1.041667~ns). The positron time is measured relative to the time of the beam particle as defined by the LV1 pulse (which is synchronised with the accelerator RF time). The analysis allows one and only one positron candidate to be found in the -8 to 22~$\mu$s window; events failing this criterion are rejected. Events are also rejected if the pion stop position falls outside an $x$ range of pixels  defined by the associated Z counter (this suppresses a small contamination of beam muons since their range is 21 pixels further than pions). After some additional geometrical and quality criteria are applied, 42\% of the LV2-triggered events are accepted for inclusion in the lifetime histograms (Figs.\,\ref{fig_lifetime}).

The appearance of muon decays at negative times involves the loss of the true positron and the gain of a fake positron. As seen in the insert of Fig.\,\ref{fig_lifetime}a, the fake positron background is structured with the accelerator RF period and dominated by incoming beam particles.  The events with negative decay times allow a precise description to be obtained for the accidental background. The events with positive lifetime are used to perform the muon lifetime measurement.

Figure \ref{fig_lifetime}b shows the same data after rebinning by exactly the accelerator RF period (``RF rebinning''), as described below, which eliminates the beam structure. A slight exponential behaviour can be seen in the background near the boundaries of the measurement window, as anticipated in the original study of systematic errors for the FAST proposal \cite{fast_proposal}. The increase in background around -8~$\mu$s is due to positrons from another pion that stops in the signal muon pixel \textit{before} the early edge of the window, whereas the increase in background around +22~$\mu$s is due to another pion that stops in the signal muon pixel (thereby faking a positron) but whose subsequent positron emerges \textit{after} the late edge of the window. In both cases the background grows as the edge of the window is approached.

\section{Muon lifetime fit}

The measurement of the muon lifetime is performed in three steps. First, the beam period is obtained with high precision from the negative decay region in Fig.\,\ref{fig_lifetime}a. Then, the lifetime histogram is rebinned using the measured beam period as the bin width (Fig.\,\ref{fig_lifetime}b). Finally a fit to the rebinned histogram is performed with the muon lifetime as one of the free parameters. The rebinning minimises the influence of the periodic background structure on the measurement of the lifetime. A systematic error due to this fit method is assessed by comparing the lifetime value obtained from the data without rebinning (\S \ref{sec_fine_bin_fit}).

Two different methods were used to obtain the beam period. In both cases, the decay interval from -7000 to -1000 TDC ticks was used, since this maximised the statistics while minimising the possible systematic uncertainties related to the exponential tail. The first method consists of folding the full interval into a single period. The dispersion of the points around the mean shape of the background is a minimum for the real beam period (\S \ref{sec_fine_bin_fit}). The second method uses the Fourier transform of the data to determine the frequency with the highest power, which corresponds to the beam frequency. Both methods agree and determine the beam period to be 18.960051 TDC ticks, with a precision of better than 1 ppm. Since this does not correspond to an integer number of TDC ticks, the data are ``RF rebinned'' by dividing fractional TDC ticks into the two adjacent bins according to a linear sharing algorithm.

A further periodicity is introduced in the data by a non-linearity of the TDCs. Although the coarse ticks are perfectly paced by the 30~MHz Rb atomic clock, the 32 fine ticks are determined by the TDC chip interpolation. This was found to have a non-linearity with an amplitude of 1 per mil, introducing periodicities of 16 and 32 ticks in the fit residuals. An identical TDC non-linearity was measured in both the beam data (by folding the residuals with a period of 1 Rb coarse clock tick) and in laboratory measurements (using random flat data). 

The positive region of the lifetime distribution in Fig.\,\ref{fig_lifetime}b is fitted with the function
\begin{eqnarray*}
N(t_{e})= f_{TDC}(t_{e})  \, ( \, A \, e^{-t_e/\tau_{\mu}} 
           +                         B \, e^{ t_e/\tau_{\mu}} 
           +                         C \, )
\end{eqnarray*}
where $t_e$ is the electron time relative to the RF bucket of the beam pion, $f_{TDC}(t_{e})$ accounts for the TDC non-linearity, and the parameters $A$, $B$ and $C$ describe the amplitude of each component. The first term represents the muon decay signal, the second term accounts for the small exponential rise of the background at the positive edge of the window, and the final term accounts for the flat, uncorrelated background. The free parameters of the fit are $A$, $B$, $C$ and $\tau_{\mu}$.

\begin{figure}[htbp]
  \begin{center}
      \makebox{\includegraphics[width=155mm]{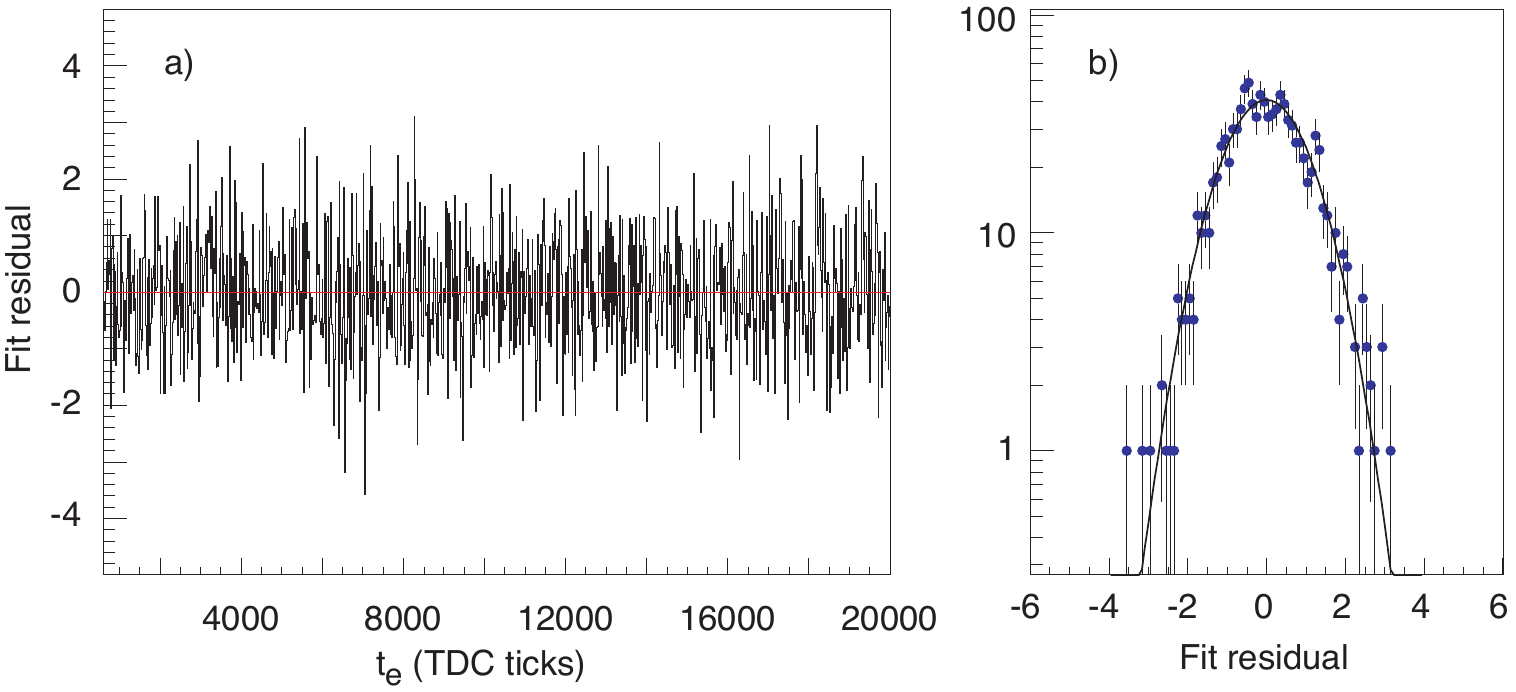}}
  \end{center}
  \caption{Residuals of the lifetime fit, defined as ($N_{data}-N_{fit})/\sqrt{N_{fit}}$: a) as a function of the decay time, and b) histogramed over the full fit interval ($\mu$: $ -0.001 \pm 0.031$, $\sigma$: $ 1.003 \pm 0.022$, $\chi^2/n_{dof}$: $ 68.3/77$, probability: 0.75).}  
  \label{fig_residuals} 
  \end{figure}

A binned maximum likelihood fit to the data is performed over the interval from 600 to 20000 TDC ticks\footnote{Since the histogram is rebinned, the actual fitting region goes from 593.8 to 20008.9 TDC ticks.}. The lower limit is chosen to avoid the ``third hit'' TDC inefficiency at earlier times, and the upper limit avoids undue sensitivity to the rising background near the edge of the time window. The result for the lifetime is $\tau_{\mu} = 2109.200 \pm 0.031$ TDC ticks [2.197 083 (32) $\mu$s], with $\chi^2/n_{dof}=1.01$ for 1020 degrees of freedom, corresponding to 40.1\% probability. 
The residuals of the fit show no systematic trends versus decay time (Fig.\,\ref{fig_residuals}a) and are Gaussian distributed, with the expected mean and width (Fig.\,\ref{fig_residuals}b). The fitted signal fraction is 96.82\%, while the positive exponential contribution is 0.02\% and the flat background accounts for 3.15\%.

Several further tests of the quality of the fit have been performed. The Fourier transform of the residuals reveals no periodicities.
Furthermore, the fit has been separated into four regions to check the stability and quality in samples that are statistically independent. The four regions are 600--3000, 3000--6000, 6000--10000, and 10000--20000 TDC ticks. Within statistical errors, the fitted lifetime is the same in all four regions, as is the fitted constant background term. As expected, the late decay region above 10000 ticks is most sensitive to the correct description of the background, while the early decays are essentially insensitive to background uncertainties. Overall, no evidence has been found to suggest any problems with the quality of the fit.

\section{Muon lifetime fit with fine binning}
\label{sec_fine_bin_fit}

An alternative analysis has been carried out using the original lifetime histogram with a fine bin size of 1 TDC tick (Fig.\,\ref{fig_lifetime}a). The data are fitted in two steps. First, a description of the shape of the background is obtained from the negative region. Second, this description plus the decay exponential and some other small corrections are used to fit the data in the positive lifetime region. The systematic uncertainties of this analysis are somewhat larger than for the nominal fit with the RF rebinned data since a more precise description is required for the DC background structure and for the pulsed, positive exponential background at the end of the TDC window. Nevertheless, this second analysis is important in order to check the consistency of the nominal fit and to assign a systematic error to the fit method. 

\begin{figure}[htbp]
  \begin{center}
      \makebox{\includegraphics[width=145mm]{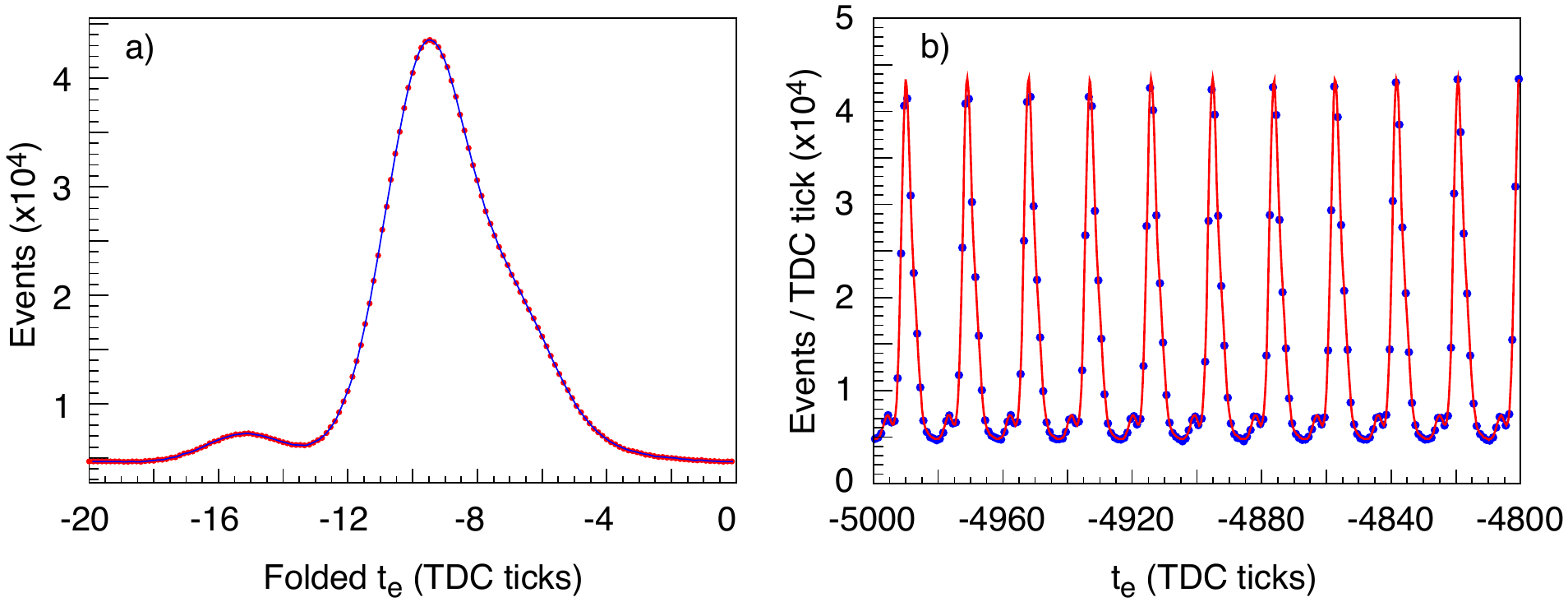}}
  \end{center}
  \caption{Spline fit to the background for the fine-binned data. a) Folded distribution showing the experimental data (curve) and the spline fit together with a constant term (points). b) Part of the negative decay region, showing the experimental data (in this case, points) together with the spline fit (curve) obtained from panel a).}  
  \label{fig_spline} 
  \end{figure}

A precise description of the negative region is obtained by folding all the data from -7000 TDC ticks to -1000 TDC ticks into one single RF period. The folded data (Fig.\,\ref{fig_spline}a) clearly show the three components of the beam background: muons (-15 TDC ticks), electrons (-9 ticks) and pions (bump near -6 ticks). The different timings between the particles are due to their different times-of-flight from the production target. The structure is well-described using a spline with 150 knots. The negative decay region is then fitted by the sum of an exponential, a flat background and the spline, repeated with the measured RF frequency. The fit---a region of which is shown in Fig.\,\ref{fig_spline}b---has a $\chi^2/n_{dof}=1.12$ for 5995 degrees of freedom. 
Once the shape of the background is fixed using the negative region, the positive lifetime region is fitted using the function:
\begin{eqnarray*}
N(t_{e})= f_{TDC}(t_e) ( A e^{-t_e/\tau_{\mu}} +
                            B e^{t_e/\tau_{\mu}} f_{\pi}(t_e) +
                            C + Spline) 
\end{eqnarray*}

\noindent where $t_e$ is the electron time relative to the beam pion, the parameters $A$, $B$ and $C$ describe the amplitude of each component, $f_{TDC}$ is the correction due to the TDC non-linearity and $f_{\pi}$ describes the shape of the increasing beam pion background near the positive edge of the window. The free parameters of the fit are $A$, $B$, $C$ and $\tau_{\mu}$.

A binned maximum likelihood fit is performed to the data between 600 and 16000 TDC ticks. The fit region excludes longer-lived muons since they are most sensitive to uncertainties in the background description. The muon lifetime obtained with these fine binned data is 5.2 ppm below the nominal value obtained with the RF rebinned data.

\section{Systematic errors}

Possible sources of systematic error have been identified and their influence on the lifetime has been evaluated in turn with the experimental data, as discussed below. 

\paragraph{1. Fit method:}

The stability of the measured lifetime is checked by varying the starting point and the end point of the fit over a large region. The results are stable and no associated systematic error can be discerned.

The influence of the algorithm for RF rebinning has been checked using three different methods: simple sharing, linear interpolation and quadratic interpolation. The observed difference between the measured lifetimes is below 0.1 ppm.

The systematic error on the TDC non-linearity correction has been estimated using two different corrections, with different amplitudes and phases. No effect is observed on the fitted lifetime.

The sensitivity to the phase of the periodic beam background has been estimated by varying this parameter over one full beam period. The associated systematic error is $\Delta \tau_{\mu}<0.1$ ppm. Furthermore the central value of the period has been varied by $\pm 100$~ppm, which is far outside the range of experimental uncertainty. The influence in the measured lifetime is $\Delta \tau_{\mu}<0.2$ ppm.

Four different methods have been used to perform the fit to test if there is any sensitivity to the analytical method: minimise the $\chi^2$, maximum the likelihood, minimise the Kullback-Leibler discrepancy and minimise the absolute value of the residuals. The various methods are sensitive to different features of the distribution. Since no differences are found, we conclude that the fitted lifetime is insensitive to the analytical method.

\begin{figure}[tbp]
  \begin{center}
      \makebox{\includegraphics[width=155mm]{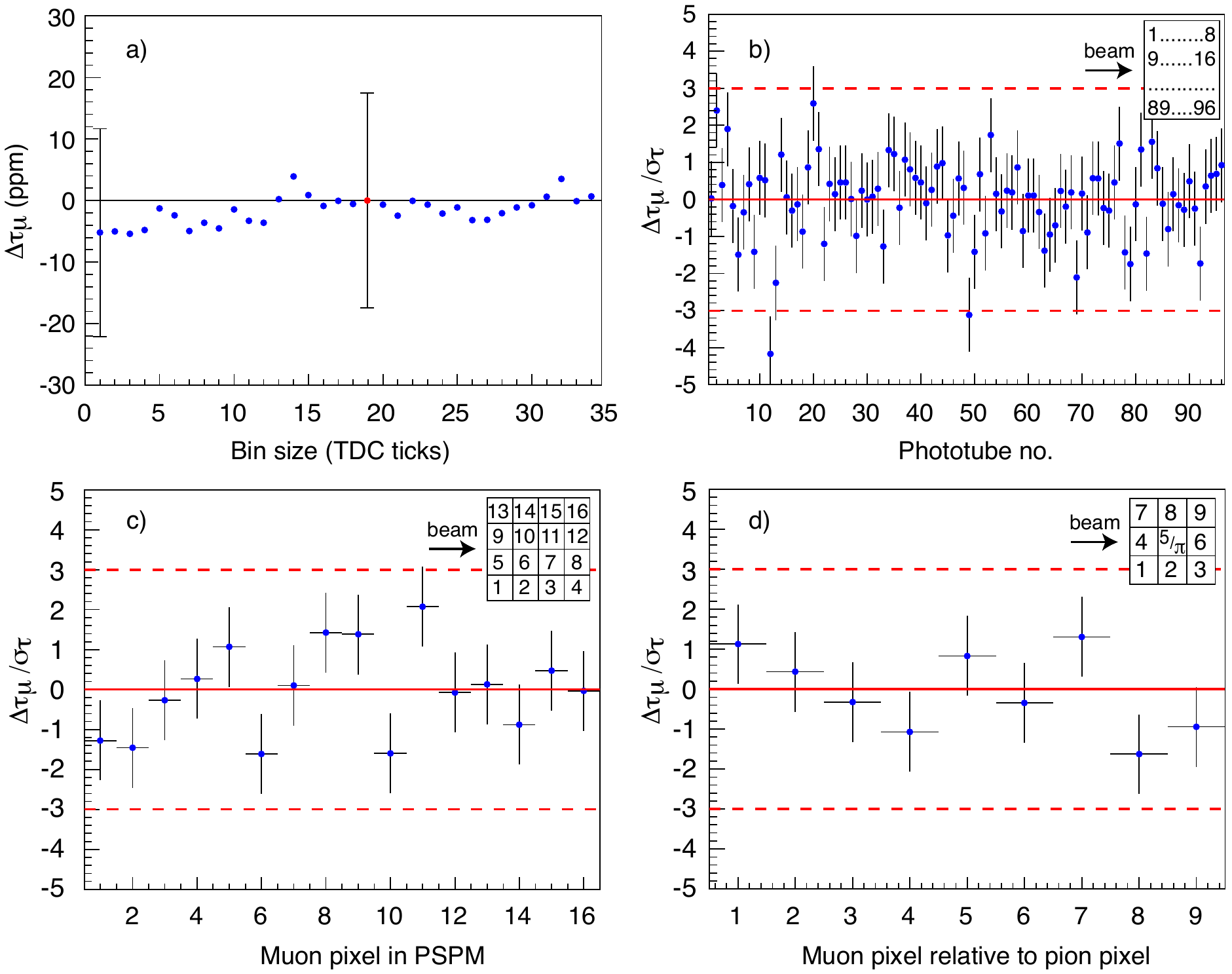}}
  \end{center}
  \vspace{-5mm}
  \caption{Examples of the evaluation of systematic errors, showing the muon lifetimes obtained as a function of various parameters: a) bin size,   b) PSPM containing the $\pi$ stop pixel,  c) $\mu$ pixel location within the PSPM, and d) position of the $\mu$ pixel relative to the $\pi$ stop pixel. In the first panel, the point at 19 TDC ticks corresponds to the nominal lifetime fit, using RF-rebinned data. The other points are derived from fits to the fine-binned data (\S \ref{sec_fine_bin_fit}). There is a systematic difference between large bins and small bins due to the required precision in the description of the background structure. In the final three panels, the lifetimes are expressed in number of standard deviations away from the nominal value. The inserts explain the numbering schemes for the horizontal axes.}  
  \label{fig_systematics} 
  \end{figure}

Finally, in order to evaluate possible biases of the rebinned method, the results have been compared with an independent measurement using the fine binned data. The method used to obtain the lifetime from the fine binned histogram is described above. The results are shown in Fig.\,\ref{fig_systematics}a for bin sizes between 1 and 34 TDC ticks. To account for the difference in the lifetime measurement between the RF rebinned method and the fine-binned (1 TDC tick) method, we ascribe a systematic error $\Delta \tau_{\mu}=-5.2$ ppm. Summing all contributions, the total systematic error associated with the fit method is therefore $-5.2$ ppm.

\paragraph{2. Reference time ($\pi$ vs.\,$\mu$):}

The nominal analysis uses the beam pion as the reference start time ($t_e = 0$), whereas the muon time could be used as an alternative. The resultant lifetime distributions are sensitive to different systematic errors.  A pion start time makes the distribution independent of the muon time and also sharpens the beam structure of the background events; a muon start time makes the analysis less sensitive to the shape of the beam background. The difference of the fitted lifetime values for the two methods ($\mu$ start minus nominal $\pi$ start) is $\Delta \tau_{\mu}=+1.8$ ppm and is taken as the associated systematic uncertainty.

\paragraph{3. Detector uniformity:}

The systematic error from detector non-uniformities is estimated as follows. First the muon lifetime is determined from independent sub-samples that divide the data according to the particular parameter under study, such as, for example, location of the muon stop pixel in the target. Any sub-samples having lifetime measurements that are statistically incompatible ($>$3$\sigma$) with the nominal lifetime measurement are then removed and a new lifetime determined from the remaining total sample.\footnote{The binning of these histograms ranges from 1 to 20 TDC ticks per bin, depending on the effect under study.} The systematic error is then estimated as the signed full difference of the new lifetime minus the nominal lifetime. 

This procedure has been used to evaluate non-uniformities with respect to the position of the pion stop pixel in the target as a function of the PSPM  (96 units), TDC module (16), TDC chip (48), $x$ coordinate  (26), and $y$ coordinate  (42). No systematic effects are observed except as a function of PSPM (Fig.\,\ref{fig_systematics}b). The measured lifetimes in two PSPMs deviate by more than $3 \sigma$ from the nominal value, and the associated systematic error is $\Delta \tau_{\mu} = +7.6$ ppm. The position of the muon decay inside the PSPM has also been checked by folding all the measured decays into one single PSPM (Fig.\,\ref{fig_systematics}c). No systematic effect is observed. The total systematic error due to target non-uniformity is therefore taken as $\Delta \tau_{\mu} = +7.6$ ppm.

\paragraph{4. Time stability:}

The stability of the experiment with respect to calendar time has been checked by  looking at the evolution of the measured lifetime during the 3 weeks over which the data taking was performed. No variation is observed.

The stability of the Rb clock has been checked against the accelerator RF clock using the Fourier transform of the negative decay time region of the lifetime plot in order to measure the accelerator RF period. The two clocks are found to have a relative stability of better than 0.5~ppm over the duration of the data taking period.

\paragraph{5. Beam rate:}

Data were recorded at LV2 trigger rates between 30 and 42 kHz. No systematic effect is observed with respect to the trigger rate.

\paragraph{6. TDC performance:}

Occasionally during data taking, a TDC chip could lose synchronisation of the fine tick phase lock. This was automatically identified when it happened by a characteristic broadening of the time distribution of the beam pion pixels. The DAQ then ended the run, declared that run as bad, re-initialised the TDCs, and started a new run. To eliminate the possibility of any residual effect from events with broadened timing, the bad runs have been collected together and fitted for the muon lifetime. The value obtained is compatible with the nominal lifetime, within the relatively large statistical errors, $\Delta \tau_{\mu}= 15 \pm 60$ ppm. Further evidence against a residual error from this source is that there was a significant difference in susceptibility to this problem among the 16 TDCs, yet there is no sign of any different lifetime measurement from the susceptible TDCs. In conclusion, there is no evidence for any systematic effect from this source.

\begin{figure}[htbp]
  \begin{center}
      \makebox{\includegraphics[width=160mm]{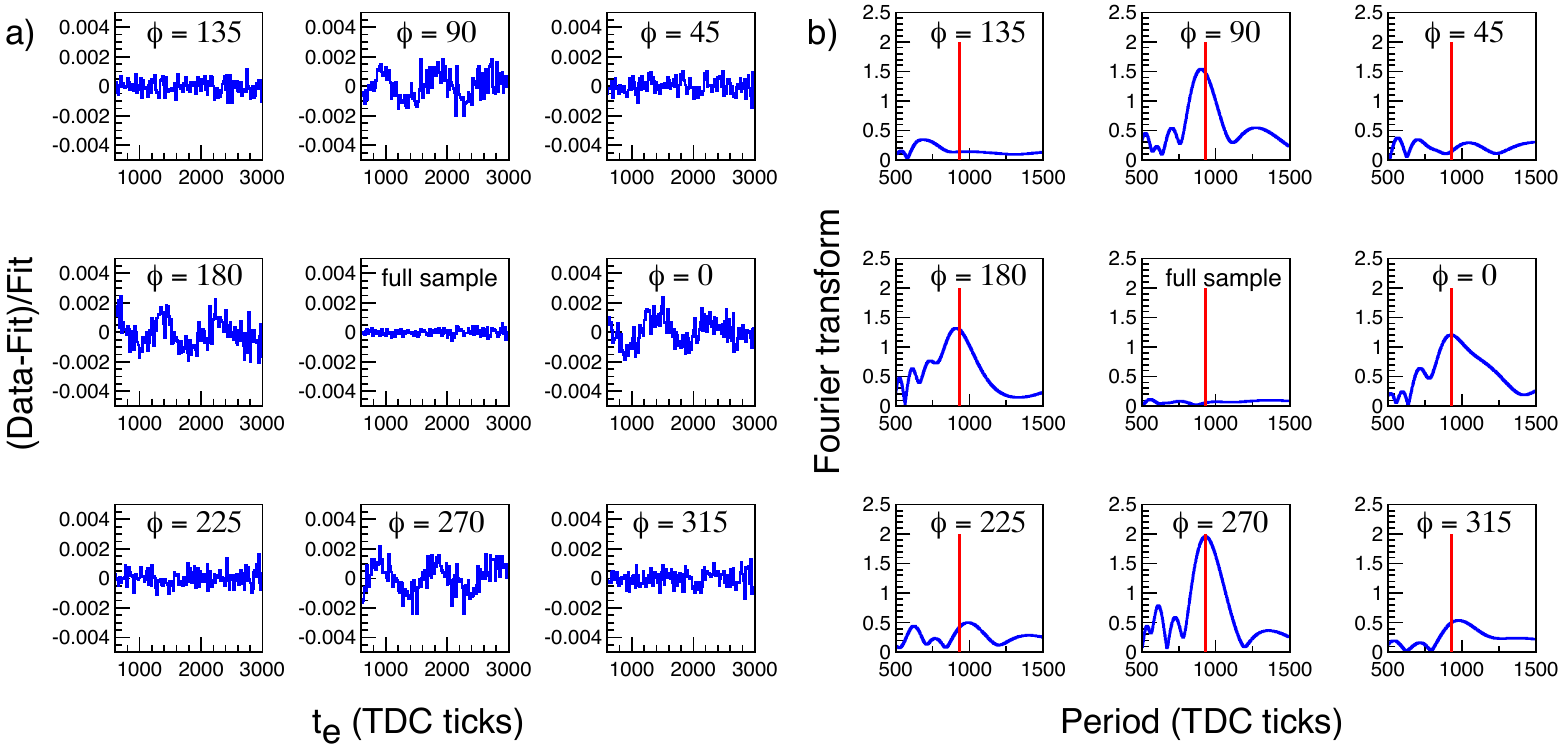}}
  \end{center}
  \caption{Study of \musr\ systematic errors. a) Fractional fit residuals with respect to the nominal lifetime distribution for positrons emerging at different angles from the muon pixel.  The fit residuals for the total sample are shown at the centre. b) Corresponding Fourier transforms of these distributions. For all abscissa, the units are TDC ticks. The \musr\ period is indicated by the vertical lines at 931 TDC ticks. Clear \musr\ signals are seen for certain positron directions, as expected, but none is present in the total sample.}  
  \label{fig_systematics_musr} 
  \end{figure}

\paragraph{7. Muon spin rotation (\musr):}

Muon spin rotation effects are highly suppressed in FAST as a result of  the basic experimental design. Nevertheless, spin effects can be extracted from the data by selection of special event topologies. These make use of the fact that the muon spin is 100\% polarised in the direction pointing back to the parent pion. A \musr\ signal can be obtained, for example, by selecting events where the $\pi$, $\mu$ and positron are in the same pixel, or where the $\pi$ and $\mu$ are in the same pixel but not the electron. From these data, the \musr\ period is measured to be 931 TDC ticks (970 ns), in agreement with that expected for $\sim$80 G magnetic field.

In order to understand \musr\ effects in the lifetime measurement, it is helpful to consider an idealised case where the pion is exactly centred on a pixel. In this case the daughter muon will lie somewhere on a sphere centred on the pion and of radius about 1.4~mm (corresponding to the range of a 4.12 MeV muon), with its spin pointing radially inwards at the time of the \pitomu\ decay. When projected onto the readout plane, this effectively produces a circular distribution of muon stop positions located (in this idealised case) near the edge of the 4~mm pixel, with spins pointing radially inwards.  The net muon spin is zero and so, assuming uniform detection efficiency, no asymmetry should be observed in the positron emission directions. The muons then begin to precess about the  magnetic field, with the largest precession angle for the forward--backward muons (i.e.\,along the beam direction and with spin direction perpendicular to the magnetic field) and the the smallest for the left--right muons (transverse to the beam direction and with spin direction along the magnetic field). 

The first check of \musr\ effects is therefore to see if there is any variation of the lifetime with respect to the relative position of the $\mu$ and the $\pi$. This is shown in Fig.\,\ref{fig_systematics}d; no systematic effects are observed.

The second check of \musr\ effects is to look at the time dependence of the residuals of the lifetime fit with respect to the direction of positron emission. This is shown in Fig.\,\ref{fig_systematics_musr}a and reveals a clear \musr\ signal in the forward--backward ($\phi=0^{\circ}$ and $180^\circ$) and left--right ($90^\circ$ and $270^\circ$) directions. The \musr\ amplitude is about 1 per mil and the phase at $t_e = 0$ gives a peak deficiency for $0^\circ$ and $180^\circ$, and a peak excess for $90^\circ$ and $270^\circ$. This is due to a small difference in the solid angle for positrons to be classified as $0^\circ$ between muons located near the upstream compared with downstream edges of the pion pixel, which arises from the relatively coarse pixel granularity. For example, at $t=0$, the downstream muons have a higher solid angle for classification as $\phi=0^\circ$ decays than do the upstream muons, which leads to a net deficiency. The mirror effect (with the same phase) occurs for positrons at $180^\circ$. The small $t=0$ deficiency at $0^\circ$ and $180^\circ$ appears as a symmetric excess at $90^\circ$ and $270^\circ$, and cancels almost completely along the diagonal directions. This pattern then oscillates with the \musr\ frequency and gradually decays as the muons depolarise. 

In short, the small \musr\ signal in the $0^\circ$ and $180^\circ$ directions due to muons emitted in the forward--backward direction is causing---and therefore exactly compensated by---an anti-phase \musr\ signal in the $90^\circ$ and $270^\circ$ directions. This is confirmed by the central distribution in Fig.\ref{fig_systematics_musr}a, which shows no \musr\ signal when all positron directions are combined. Figure \ref{fig_systematics_musr}b shows the corresponding Fourier transform for each of the positron directions. The \musr\ signal is seen to be completely absent in the summed data. We conclude that no systematic errors are present due to \musr\ effects. Furthermore, the equality of the \musr\ signal in the $0^\circ$ and $180^\circ$ directions confirms that there is a negligible residual component of beam muons, which would produce a forward--backward \musr\ asymmetry. 

\begin{table}[htbp]
\begin{center}
\caption{Summary of the systematic errors in the muon lifetime measurement. The total systematic error is obtained by first summing the same-sign errors in quadrature and then taking the average of the absolute values of the positive and negative errors.}
\label{tab_systematics}
  \vspace{2ex}
\begin{tabular}{r l r}
\hline\hline
\multicolumn{2}{l}{\textbf{Source of systematic error}}  & \boldmath $\Delta \tau_\mu$ \\
 &    & \textbf{(ppm)} \\
\hline
1. & Fit method                              & -5.2    \\
2. & Reference time ($\pi$ vs.\,$\mu$)       & +1.8    \\
3. & Detector uniformity       	  		   	 & +7.6    \\
4. & Time stability                          & $< 1  $ \\
5. & Beam rate                               & $< 1  $ \\
6. & TDC performance                         & $< 1  $ \\
7. & Muon spin rotation	   					& $< 1  $ \\
\hline
& {\bf Total}                  & \mbox{\boldmath $\pm 6.5$} \\
\hline\hline
\end{tabular}
\end{center}
\end{table}

\section{Final result and conclusion}

A summary of the estimated contributions from the various sources of systematic uncertainty is shown in Table~\ref{tab_systematics}, leading to a total systematic error of 6.5~ppm.  The measured value of the \mupl\ lifetime is therefore 2109.200 $\pm$ 0.031 $\pm$ 0.014 TDC ticks, i.e.
\begin{eqnarray*}
\tau_\mu & = & 2.197 \; 083 \; (32) \; (15) \; \mu \textrm{s}, 
\end{eqnarray*}
where the first error is statistical and the second is systematic.
This measurement, with 16~ppm overall precision, is in good agreement with both the previous world average (18~ppm precision) \cite{pdg06} and the new result from MuLan (11~ppm precision) \cite{mulan07}. From Eq.\,\ref{eq_gf}, our measurement of the muon lifetime determines the Fermi constant to be
\begin{eqnarray*}
G_\textrm{\scriptsize F} & = & 1.166 \; 353 \; (9)   \times 10^{-5}  \; \textrm{GeV}^{-2}, 
\end{eqnarray*}
which may be compared to the world average value of $1.166 \; 370 \; (10) \times 10^{-5}$ GeV$^{-2}$ in the 2006 Review of Particle Properties \cite{pdg06}.

The uncertainty of our present measurement of \gf\ to 8 ppm precision is dominated by the statistical error. The size of the data sample also limits the determination of the systematic errors. It is expected that both will decrease with larger data samples, allowing the FAST experiment to reach its final goal of determining the Fermi constant \gf\ to 1~ppm precision.

\newpage

\section*{Acknowledgements}

We would like to acknowledge and thank several colleagues who were important in the design studies and early preparations of the experiment:  P.\,de Jong, D.\,della Volpe, P.\,Kammel, R.\,Nahnhauer, F.\,Navarria, G.\,Passaleva, A.\,Perrotta, R.\,Stuart and G.\,Valenti. We also thank A.\,Dijksman, D.\,George and T.\,Rauber for their important contributions to the FAST hardware. Finally we acknowledge the generous support of PSI, and the PSI accelerator groups for their efficient operation of the cyclotron and beamline.

\end{document}